\documentstyle[12pt]{article}
\def\BS{{\cal B}}
\def\dl{\delta }
\def\DS{{\cal D}}
\def\HS{{\cal H}}
\def\PS{{\cal P}}

\def\gm{\gamma }
\def\Gm{\Gamma }
\def\Pt{{\tilde P}}
\def\sg{\sigma }
\def\th{\theta }

\def\beq{\begin{equation}}
\def\beqa{\begin{eqnarray}}
\def\eeq{\end{equation}}
\def\eeqa{\end{eqnarray}}
\def\hf{{\textstyle {1\over2}}}

\input{epsf}
\begin{document}

\title{Consistent histories, quantum truth functionals, and hidden variables}

\author{Robert B. Griffiths\thanks{Electronic mail: rgrif@cmu.edu}\\
Department of Physics,
Carnegie-Mellon University,\\
Pittsburgh, PA 15213, USA}

\date{Version of 8 Sept. 1999}

\maketitle

\begin{abstract}
	A central principle of consistent histories quantum theory, the
requirement that quantum descriptions be based upon a single framework (or
family), is employed to show that there is no conflict between consistent
histories and a no-hidden-variables theorem of Bell, and Kochen and Specker,
contrary to a recent claim by Bassi and Ghirardi.  The argument makes use
of  ``truth functionals'' defined on a Boolean algebra of classical or quantum
properties. 
\end{abstract}

\section{Introduction}
\label{s1}

	Textbook quantum theory is beset by numerous conceptual difficulties
when it comes to providing a physical interpretation of the mathematical
formalism of quantum theory. The unsatisfactory nature of the usual treatments
results in a variety of paradoxes: Schr\"odinger's cat,
Einstein-Podolsky-Rosen, and the like
\cite{wz83}.  The consistent histories (sometimes called decoherent histories) 
approach \cite{gr84,om88,gmh90} to quantum mechanics disposes of these
difficulties by combining probability theory with the standard Hilbert space
formalism in a coherent way through the use of {\it histories} which satisfy
certain {\it consistency} conditions.

	A central principle of the more recent formulations of consistent
history (CH) ideas \cite{om94,gr96,gr98,om99} is the {\it single framework}
(also known as the single family, single logic, or single set) rule, which
plays an essential role in ensuring the logical consistency of quantum theory,
resolving various quantum paradoxes, and getting rid of the mysterious
superluminal influences which are sometimes thought to be a consequence of the
quantum formalism.  The purpose of this Letter is to explain how the single
framework rule renders the CH approach fully compatible with a well-known
result on the impossibility of hidden variables in a Hilbert space of dimension
three or greater due to Bell \cite{bl66}, and Kochen and Specker \cite{ks67}.
Following recent work by Bassi and Ghirardi \cite{bg99}, we shall explore the
problem using the notion of a {\it truth functional}.  While this is not
absolutely essential---the ideas of ordinary probability theory suffice, when
they are properly used---it is convenient for the conceptual issues we will be
discussing, and assists in explaining why our key conclusions are directly
contrary to those of \cite{bg99}.

	Since we will be concerned with the Hilbert space description of a
system at a single time, most of the formal machinery of the CH
approach---histories, consistency, and assignment of probabilities---is not
needed for the following discussion.  This will allow us to concentrate on the
most essential point, which is that a quantum Hilbert space differs in crucial
respects from a classical phase space, and this mathematical difference must be
reflected in a physical interpretation of the theory.  The idea of a truth
functional, which may be unfamiliar to some readers, is developed in
Sec.~\ref{s2} using a classical phase space, where classical intuition is a
reliable guide, and then applied in Sec.~\ref{s3} to a quantum Hilbert space,
where certain classical ideas run into difficulty.  It is at this point that
results on no-hidden-variables are used, illustrated with a two-spin paradox
due to Mermin \cite{mr93}, based on an idea of Peres \cite{pr90}. The CH
approach to truth functionals and the no-hidden-variables theorem is the
subject of Sec.~\ref{s4}, while the claims found in \cite{bg99} are discussed
in Sec.~\ref{s5}.

\section{Classical Truth Functionals}
\label{s2}

	Consider a classical mechanical system, such as a simple harmonic
oscillator, described by a phase space $\Gm$, with $\gm$ a representative
point.  Any physical property $P$ of the system corresponds to some set of
points $\PS$ in the phase space for which this property is true, and we shall
define the corresponding {\it indicator} function $P(\gm)$ to be 1 whenever
$\gm$ is in $\PS$, and 0 otherwise.  For example, let $P$ be the property that
the total energy of the oscillator is less than some constant $E_0$.  Then
$\PS$ is the the region inside an ellipse in the $x,p$ plane ($x$ the position,
$p$ the momentum), and $P(\gm)$ is 1 for $\gm$ inside and 0 for $\gm$ outside
this ellipse. If $I$ is the function equal to 1 everywhere on the phase space,
the indicator of the negation $\Pt$ of $P$, ``energy greater than or equal to
$E_0$'' in our example, is the function $I-P$: it is 0 wherever $P$ is 1, and 1
wherever $P$ is 0.  Likewise, if $P$ and $Q$ are any two properties, the
product $PQ$ of the two indicators is the indicator for the conjunction $P\land
Q$ of $P$ and $Q$, the property ``$P$ AND $Q$''.  Similarly, the disjunction
$P\lor Q$, ``$P$ OR $Q$'', corresponds to the indicator $P+Q-PQ$.

	Consider a coarse graining of the phase space into a collection
$\DS$ of $N$ non-overlapping regions or ``cells''.  Then we can write
\beq
  I=\sum^N_{j=1} D_j,
\label{e1}
\eeq
where $D_j$ is the indicator corresponding to the $j$'th cell. Since the  cells
do not overlap it follows that
\beq
  D_j D_k = \dl_{jk} D_j, 
\label{e2}
\eeq
consistent with the obvious fact that $I^2 = I$. 
The set of
$2^N$  indicators of the form
\beq
  P=\sum^N_{j=1} \pi_j D_j,
\label{e3}
\eeq
where $\pi_j$ is either 0 or 1, form a Boolean algebra $\BS$ of properties, in
which $P\cap Q$ is $P\land Q$, and $P\cup Q$ is $P\lor Q$, as defined above. 

	Given such a Boolean algebra $\BS$, we define a {\it truth functional}
$\th$ to be a function which assigns to every property $P$ in $\BS$ the value 1
(true) or 0 (false) in a way which satisfies the following three conditions:
\beq
  \th(I)=1,\quad \th(I-P) = 1-\th(P),\quad \th(PQ) = \th(P) \th(Q).
\label{e4}
\eeq
These correspond to the rather sensible requirements that something is always
true, that $P$ is true if and only if its negation $\Pt = I-P$ is false, and
both $P$ and $Q$ are true if and only if their conjuction $P\land Q$ is true. 

	 One can show that for a given coarse graining $\DS$ with Boolean
algebra $\BS$, there is a one-to-one correspondence between truth functionals
on $\BS$ and the elements of $\DS$.  That is to say, any function $\th$ taking
only the values 0 and 1 and satisfying (\ref{e4}) must be of the form
\beq
  \th_k(P) = \cases { 	1 & if $PD_k = D_k$,\cr
			0 & if $PD_k = 0$.}
\label{e5}
\eeq
for some $k$.  In terms familiar from probability theory, one can regard the
non-overlapping cells which constitute the coarse graining $\DS$ as a {\it
sample space} of mutually exclusive possibilities, one and only one of which
actually occurs, or is ``true'', namely the cell which contains the phase point
$\gm$ which represents the actual or true state of the system.  From this
perspective $\th_k(P)$ is the probability of the property $P$ conditional upon
the property $D_k$, and we have the usual identification of ``true'' with
``probability one'' and ``false'' with ``probability zero''.

	Notice that it is because we are assuming that $P$ is of the form
(\ref{e3}) that the product $PD_k$ must have one of the two forms on the right
side of (\ref{e5}): no property of the form (\ref{e3}) can include part but not
all of some cell $D_k$.  Consequently, (\ref{e5}) defines a truth functional
for indicators belonging to this particular algebra $\BS$, but not for all
possible properties; in this sense a truth functional is relative to a
particular decomposition $\DS$, or Boolean algebra $\BS$. However, it is also
possible to construct a {\it universal truth functional} which is not limited
to a single Boolean algebra, but which will assign 0 or 1 to {\it any}
indicator on the classical phase space in a manner which satisfies (\ref{e4}).
To do this, choose some point $\gm_0$ in $\Gm$, and let
\beq
  \th_0(P)=P(\gm_0).
\label{e6}
\eeq
That is, $\th_0$ assigns the value 1 to any property which contains the point
$\gm_0$, and 0 to any property which does not contain this point, in agreement
with how one would normally understand ``true'' in a case in which the state of
the system is correctly described by $\gm_0$.  As we shall see, a key
difference between classical and quantum physics is the fact that in the latter
there are no universal truth functionals as long as the Hilbert space has a
dimension greater than 2.

\section{Quantum Truth Functionals}
\label{s3}

	The quantum counterpart of a classical phase space is a Hilbert space
$\HS$.  For our purposes it suffices to consider cases in which $\HS$ is of
finite dimension, thus avoiding the mathematical complications of
infinite-dimensional spaces.  The counterpart of a classical property is a
linear subspace $\PS$ of $\HS$, with a corresponding orthogonal projection
operator or {\it projector} $P$.  If $I$ is the identity operator, the negation
$\Pt$ of the property $P$ corresponds to the projector $I-P$, and the
conjunction $P\land Q$ of two properties corresponds to the projector $PQ$ {\it
in the case in which $P$ and $Q$ commute with each other}.  If $PQ\neq QP$,
then neither $PQ$ nor $QP$ is a projector, so there is no obvious way to define
a property corresponding to the conjunction.  We shall return to this later, in
Sec.~\ref{s4}.

	The quantum counterpart of a coarse graining of a classical phase space
is a {\it decomposition $\DS$ of the identity} which has precisely the form
(\ref{e1}), where the $D_j$ are now mutually orthogonal projectors which
satisfy (\ref{e2}).  This decomposition gives rise to a set of projectors of
the form (\ref{e3}), all of which commute with each other, and which form a
Boolean algebra $\BS$ analogous to the algebra of classical indicator
functions.

	A quantum truth functional $\th$ assigns to every projector $P$ in the
Boolean algebra (\ref{e3}) the value 0 or 1 in a way which satisfies the three
conditions in (\ref{e4}).  Once again, there is a one-to-one correspondence
between truth functionals and the elements of $\DS$, and each such truth
functional is of the form (\ref{e5}) for some $k$.  The intuitive
interpretation is also similar in the quantum and classical cases.  The
projectors which enter the decomposition $\DS$, or equivalently the
corresponding subspaces of $\HS$, form a sample space of mutually exclusive
possibilities, one and only one of which is a correct description of the
system, and thus ``true''.  Larger subspaces in $\BS$ which contain this true
space are also true, and the others are false.

	A truth functional associated with the decomposition $\DS$ can be used
to assign a (real) numerical value to an observable corresponding to a
Hermitian operator $A$ written in the form
\beq
  A=\sum_j a_j D_j,
\label{e7}
\eeq
where the $a_j$ are, of course, the eigenvalues of $A$.  Thus if the truth
functional is $\th_k$, (\ref{e5}), then $D_k$ is true, and any $D_j$ with $j$
unequal to $k$ is false, so $\th_k$ assigns to $A$ the (eigen)value $a_k$.
Similarly, given some collection of observables represented by {\it commuting}
Hermitian operators $A, B, C,\ldots$, the operators can be simultaneously
diagonalized using a single decomposition of the identity.  Then a truth
functional associated with this decomposition will assign numerical values to
each of the operators in a consistent way, so that, for example, if $A$, $B$,
and $C$ are assigned the eigenvalues $a$, $b$, and $c$, then, for example, the
operator $AB + 2C$---which, of course, can also be represented using the same
decomposition of the identity---will be assigned the value $ab+2c$.

	In analogy with the classical case, Sec.~\ref{s2}, let us define a
{\it universal quantum truth functional} $\th$ as one which assigns to {\it
every} projector $P$ (thus every subspace) of $\HS$ one of the two values 0 or
1 in a way which satisfies the rules in (\ref{e4}), with, however, the
following qualification.  If two projectors $P$ and $Q$ do not commute, so that
$PQ$ is not a projector, then the third rule in (\ref{e4}) should be ignored;
we only require that it hold in cases in which $PQ=QP$. When applied to
projectors belonging to the Boolean algebra $\BS$ associated with some
particular decomposition $\DS$ of the identity, a universal truth functional
has the same properties as an ``ordinary'' truth functional; that is, if the
decomposition is (\ref{e1}), then $\th$ coincides, on this Boolean algebra,
with $\th_k$, (\ref{e5}), for some specific $k$.  Consequently, a universal
truth functional, if it exists, can be used to assign to {\it every} observable
(Hermitian operator) on $\HS$ one of its eigenvalues, and for commuting
collections of Hermitian operators these values will satisfy the usual
algebraic rules associated with ordinary numbers when one considers products
and sums of operators, as in the example considered above.

	Alas, it was shown by Bell \cite{bl66}, and by Kochen and Specker
\cite{ks67} that if $\HS$ has a dimension of three or more,
universal truth functionals do not exist.  A very simple nonexistence proof for
a Hilbert space of dimension 4 is provided by Mermin's paradox for two
spin-half particles \cite{mr93,pr90}, based upon the following ``magic
square'', which was also used in \cite{bg99}:
\beq
  \matrix{		&		  &		    \cr
		\sg^a_x	&  \sg^b_x        & \sg^a_x\sg^b_x  \cr
			&		  &		    \cr
		\sg^b_y	&  \sg^a_y        & \sg^a_y\sg^b_y  \cr
			&		  &		    \cr
	 \sg^a_x\sg^b_y & \sg^a_y\sg^b_x  & \sg^a_z\sg^b_z  \cr
			&		  &			}
\label{e8}
\eeq
Here $\sg^a_x$ is the Pauli $\sg_x$ operator for spin $a$, $\sg^b_y$ the
$\sg_y$ operator for spin $b$, and so forth.  Note that the operators for
particle $a$ commute with those for particle $b$, whereas the commutator of
$\sg^a_x$ and $\sg^a_y$ is $2i\sg^a_z$, etc.  Each of the nine operators in the
square has eigenvalues $+1$ and $-1$, and each eigenvalue is two-fold
degenerate.  The three operators in each row in (\ref{e8}) commute with each
other, as do the three operators in each column.  In addition, it is not hard
to show that the product of the three operators in each row is the identity
$I$.  The product of the three operators in each of the first two columns is
$I$, but the product of those in the third column is $-I$.

	These mathematical properties are incompatible with the existence of a
universal truth functional.  For suppose that such a functional existed.  Then,
as explained earlier, it could be used to assign a numerical eigenvalue of
$\pm 1$ to each of the nine operators in the square.  Since the operators in
each row commute with one another, the usual algebraic properties would be
preserved for the numerical assignments corresponding to this row.  This would
mean that the product of the numbers in any given row would have to be
$+1$, since a truth functional must assign to $I$ its only eigenvalue, $+1$.
Similarly, the product of the numerical values in the first two columns would
have to be $+1$, and in the last column it would have to be $-1$.  But no
such assignment of numerical values exists.  For example,
\beq
  \matrix {	-1  &  -1 &  +1  \cr
		+1  &  -1 &  -1  \cr
		-1  &  -1 &  +1, }
\label{e9}
\eeq 
satisfies all the product rules, except that the product of the values in the
second column is $-1$, not $+1$.  To see that {\it no} assignment of $\pm 1$
can satisfy all the rules, find the product of the three numbers in every row,
next the product of the three numbers in every column, and, finally, take the
product of all six of these products.  The result will be the product of the
squares of all nine entries in the $3\times 3$ matrix, thus 1, whereas the
rules would require that it be $-1$.

\section{Consistent Histories and Truth Functionals}
\label{s4}

	How does the consistent histories approach deal with Mermin's magic
square and similar paradoxes?  To understand this, let us return to a problem
mentioned earlier, that of making sense of the conjunction $P\land Q$, ``$P$
AND $Q$'', of two quantum properties when the projectors do not commute,
$PQ\neq QP$.  (Notice that this problem never arises in classical physics,
since the product of two indicators on the classical phase space is the same in
either order.)  For example, for a spin-half particle, the projector for the
property $\sg_y=+1$ is $\hf (I+\sg_y)$, and that for the property $\sg_x=+1$ is
$\hf (I+\sg_x)$.  As these projectors obviously do not commute with each other,
can one make sense of the statement $\sg_y=+1$ AND $\sg_x=+1$?

	The answer of the consistent historian is that one {\it cannot} make
sense of $\sg_y=+1$ AND $\sg_x=+1$: it is a meaningless statement in the sense
that the CH interpretation assigns it no meaning.  In the CH approach there are
no hidden variables, and thus there is a one-to-one correspondence between
quantum properties and subspaces of the Hilbert space.  Since every
one-dimensional subspace of the two-dimensional Hilbert space $\HS$ of a
spin-half particle corresponds to a spin in a particular direction, there is
none left over which could plausibly represent $\sg_y=+1$ AND $\sg_x=+1$.

	To be sure, one might consider assigning to $\sg_y=+1$ AND $\sg_x=+1$
the zero element of $\HS$, which is a zero-dimensional subspace corresponding
to the property which is always false, analogous to an indicator which is
everywhere zero on a classical phase space. This, in fact, was the proposal
(for this situation) of Birkhoff and von Neumann in their discussion of quantum
logic \cite{bvn36}.  It is important to notice the difference between their
approach and the one used in CH.  A proposition which is meaningful but false
is very different from a meaningless proposition: the negation of a false
proposition is a true proposition, whereas the negation of a meaningless
proposition is equally meaningless.  The Birkhoff and von Neumann approach
requires, as they pointed out, a modification of the ordinary rules of
propositional logic, whereas the CH approach does not.  However, in CH quantum
theory it is necessary to exclude meaningless properties from meaningful
discussions, which is not a trivial task.

	In particular, in CH quantum theory any quantum
description of a single system at a particular time must employ a {\it single
framework}, which is to say a single Boolean algebra of commuting projectors
based upon a definite decomposition of the identity.  To be sure, alternative
descriptions can be constructed using different decompositions of the identity,
but these cannot be combined to form a single description, nor can logical
reasoning about a quantum system be carried out by combining results from two
different Boolean algebras.  This is the {\it single framework rule}, which is
central to a correct understanding of CH quantum theory, and the point most
often misunderstood by physicists unfamiliar with the CH approach.   

  	In some cases results for two different Boolean algebras can be
combined by using the device of a ``common refinement'': a third algebra which
includes all the projectors of the first two algebras.  If a common refinement
exists, the two algebras (or frameworks) are said to be {\it compatible}; if
not, they are {\it incompatible}. If two algebras are compatible, one can use
the common refinement instead of the original algebras as the single framework
required by the single framework rule.  However, such a refinement exists if
and only if all the projectors in one of the algebras commutes with all of the
projectors in the other algebra.  Consequently, it is not possible to produce a
consistent quantum description which combines results from two Boolean algebras
when some of the projectors in one do not commute with some of the projectors
in the other.

	The single framework rule as applied to Boolean algebras of properties
refers to a {\it single system} at a {\it single instant} of time. Given two
nominally identical systems, there is no reason why one cannot use one
framework for the first and a different framework for the second.  For
instance, in the case of the two spin-half particles in (\ref{e8}), there is no
problem using $\sg_x$ for one and $\sg_y$ for the other.  Thus, whereas
$\sg_y=+1$ AND $\sg_x=+1$ is a meaningless expression for a single particle,
$\sg^a_y=+1$ AND $\sg^b_x=+1$ makes perfectly good sense. Conversely, when
incompatible frameworks turn up in some quantum discussion, it is best to think
of them as referring to two different systems, or to a single system at two
different times. (The latter is an example of a history, and when histories
involve three or more times, the CH approach imposes additional rules---but
these are outside the scope of the present discussion.) 

	In view of the preceding remarks, the reader will not be surprised to
learn that truth functionals are meaningful constructions within CH quantum
theory {\it provided} they refer to a {\it single} framework or Boolean
algebra.  Thus a {\it universal} truth functional makes no sense as soon as the
Hilbert space is of dimension two, since one already has projectors which do
not commute with each other, and simultaneously (i.e., with a single truth
functional) assigning truth values to properties which cannot simultaneously
enter the same quantum description is meaningless.  The same objection applies
to universal truth functionals in higher-dimensional Hilbert spaces, but, as
already noted, there are no such things in Hilbert spaces of dimension three
or more!

	As the single framework rule may seem a bit abstract, let us see how it
applies to the case of the magic square (\ref{e8}).  As long as one considers a
single row, the operators commute with each other, and hence they are of the
general form (\ref{e7}) using a {\it single} decomposition of the identity.
There is, therefore, no problem with introducing a truth functional which
assigns to each of these operators one of its eigenvalues; for example, those
in the top row of (\ref{e9}). Of course, the same remark applies to the
operators in the first column of (\ref{e8}): they commute with each other, and
can all be expressed in terms of a single decomposition of the identity.
However, this decomposition of the identity is {\it different} from the one
used for the first row, and the two are incompatible---they have no common
refinement---as is immediately obvious from the fact that $\sg^b_x$ in the
first row does not commute with $\sg^b_y$ in the first column.  Consequently,
it makes no sense to {\it simultaneously} assign values to all the operators in
the top row {\it and} those in the first column, much less values to all nine
operators in the square.  For this reason it is impossible to construct a
paradox if one pays attention to the CH rules.

\section{The Argument of Bassi and Ghirardi}
\label{s5}

	In \cite{bg99} Bassi and Ghirardi use the magic square (\ref{e8}) to
argue that CH ideas when combined with three assumptions which they regard as
reasonable, and even necessary for a sound interpretation, when combined with a
fourth assumption usually made by proponents of the CH interpretation (but
which they themselves find questionable) lead to a logical contradiction. The
four assumptions can be stated briefly as follows, where the language has been
changed so that it refers to a system at a single time (rather than to
histories), as this is all that is needed for the present discussion:
\begin{description}
\item[(a)] Ordinary rules of classical reasoning apply to a single Boolean
algebra of projectors. 
\item[(b)] It is possible to assign a truth value
to every projector which occurs in a particular Boolean algebra. 
\item[(c)] A projector should be assigned the same truth value in all Boolean
algebras which contain it. 
\item[(d)] Any Boolean algebra of projectors can be used to construct a
legitimate quantum description.
\end{description}
From these assumptions Bassi and Ghirardi deduce the existence of a universal
truth functional, to use the terminology employed in this Letter, and then show
that such an object is inconsistent with the mathematical properties of the
operators in (\ref{e8}).

	Assumptions (a) and (d) are part of the standard CH approach.  As for
(b), we noted in Sec.~\ref{s3} that if a decomposition of the identity contains
$N$ projectors, there are $N$ distinct truth functionals associated with the
corresponding Boolean algebra $\BS$, so it is always possible to choose one of
them and use it to assign a truth value to the projectors in $\BS$.  Thus (b)%
\footnote{ A statement in Sec.~5 of \cite{bg99}, could be interpreted to mean
that (b) is equivalent to the assumption that {\it every} property has a truth
value; however the interpretation given here coincides with that intended by
the authors \cite{bgpc}.}
agrees with the CH interpretation.

	Assumption (c) is ambiguous.  As noted in Sec.~\ref{s3}, any quantum
truth functional is associated with a particular Boolean algebra.  One could
understand (c) to mean that for a {\it given} projector $P$, we should consider
all Boolean algebras which contain it, and in each of these algebras we should
pick a truth functional which assigns the same value, say 1, to $P$.  This can
certainly be done, as a mathematical exercise, for a particular projector $P$.
However, the authors of \cite{bg99} mean something quite a bit stronger
\cite{bgpc}.  Namely, that {\it every} projector on $\HS$---or, if one does not
make assumption (d), every projector belong to some collection of acceptable
properties---is assigned a definite truth value, 1 or 0, independent of the
Boolean algebra which contains it.  Understood in this way, (c) combined with
(a), (b), and (d) is equivalent to the assumption of a universal truth
functional, an impossibility for a Hilbert space of dimension 3 or more, as
explained in Sec.~3, and contrary to CH quantum mechanics for the reasons
indicated in Sec.~4.

	However, the single framework rule of CH quantum theory is already
inconsistent with the first (weaker) interpretation of (c) in the preceding
paragraph, in the following sense.  Suppose that $P$ belongs to two
incompatible Boolean algebras $\BS'$ and $\BS''$ containing projectors which do
not commute with each other.  Then $\BS'$ and $\BS''$ cannot be combined in a
single quantum description. Therefore, if both are to be employed, they cannot
refer to the same physical system at the same time.  But if we are dealing with
two physical systems, or the same system at two different times, there is, in
general, no reason to suppose that two truth functionals should coincide for
$P$ or for any other projector in $\BS'\cap\BS''$.

	It may help to consider a specific example.  Let $\BS'$ be the Boolean
algebra corresponding to the operators on the first row of the magic square
(\ref{e8}), $\BS''$ that of the first column, and choose truth functionals
$\th'$ and $\th''$ which assign (as discussed in Sec.~\ref{s3}) the values
shown in the first row and the first column of (\ref{e9}), respectively.  Then
both $\th'$ and $\th''$ assign to $\sg^a_x$ the value $-1$, in accordance with
the weaker interpretation of (c).  Nonetheless, they cannot possibly refer to
the same physical situation, because $\th'$ assigns to $\sg^b_x$ the value
$-1$, and $\th''$ to $\sg^b_y$ the value $+1$.  But a single system at a single
time cannot have both a value for $\sg^b_x$ {\it and} a value for $\sg^b_y$ if
one uses the Hilbert space of standard quantum mechanics, for the reasons
discussed in Sec.~4 above. Consequently, the fact that $\th'$ and $\th''$ assign
the same value, $-1$, to $\sg^a_x$ is, from the CH point of view, devoid of any
particular significance, since if these truth functionals are describing
different systems this agreement is fortuitous. 

  	In summary, if (c) is understood in the stronger of the two senses
discussed above it leads, in combination with (d), to the existence of a
universal truth functional, which conflicts with the Bell-Kochen-Specker
results as well as with consistent histories.  However, there is already a
clear conflict with the single framework rule of CH quantum theory as soon as
(c) is interpreted in a weaker way which only requires using two or more
incompatible Boolean algebras to refer to a single physical system at the same
time.

	Unfortunately, the single framework rule, despite the prominence given
to it in several of the references cited in their Letter, is not mentioned
by Bassi and Ghirardi  when they introduce what they consider to be the basic
principles of CH quantum theory, nor is it referred to, except somewhat
obliquely in their Sec.~2(c), until after they have completed their definitions
and their main argument. When at the beginning of Sec.~4 they state the rule in
its entirety for the first time, they admit that their argument is, indeed, in
conflict with this rule, but then offer the excuse that they are employing a
different form of reasoning from that employed in CH quantum theory. While the
single framework rule is appropriate for the latter, it cannot, they assert,
apply to the former.  

	There seems to be no reason to debate this point, as Bassi and Ghirardi
are surely not obligated to adopt the rules for quantum reasoning which the
developers of the CH approach regard as most appropriate.  One could wish,
however, that they had made plain much earlier in their Letter that it is not
``standard'' consistent histories quantum theory, but instead an alternative
version they themselves invented, with different rules of reasoning, which
leads to a logical contradiction.  In this regard theirs can be added to a list
of work by other authors%
\footnote{See Sec.~V C and D, and App.~A of \cite{gr98}; 
also \cite{ghk98,sh98}.}\ 
which shows that attempting to
construct a histories interpretation of quantum theory while omitting the single
framework rule generally leads to unsatisfactory results.

	Although the claim that (standard) CH leads to a logical contradiction
is unfounded for the reasons just noted, there is another aspect of \cite{bg99}
which deserves comment.  Translated into the language used in this Letter, the
authors make what is, in essence, the claim that one cannot treat quantum
properties as part of an ``objective reality'' if one denies the existence of a
universal truth functional.  This is the sort of conceptual and philosophical
issue which cannot be settled by an appeal to logic and mathematics; instead it
requires the application of physical intuition and an exercise of judgment.  It
is the case that in CH quantum theory, properties at a single time, and
histories, which are sequences of quantum properties at successive times, are,
under appropriate conditions, considered to be ``real'' and ``objective''.
Some issues involved in treating the CH approach as a realistic interpretation
of quantum theory are discussed in Sec.~V~B of \cite{gr98}, and the conclusion,
translated into the terminology of the present Letter, is that there is no
reason why one should regard the existence of universal truth functionals as a
necessary part of quantum reality.  Rather than repeat the argument here, let
me simply note that I believe that \cite{gr98} provides a quite adequate
response, even though it was written earlier, to the concerns raised in
\cite{bg99}.

\section*{Acknowledgments}

	 The author is indebted to T. A. Brun and O. Cohen for reading the
manuscript, and to A. Bassi and G. C. Ghirardi for correspondence regarding
\cite{bg99}.   The research described here was supported by the National Science
Foundation Grant PHY 99-00755

\end{document}